\documentclass[aps,groupedaddress]{revtex4}
\usepackage{epsfig}
\usepackage{graphicx}
\usepackage[T1]{fontenc}
\usepackage{ae}
\usepackage{color}
\usepackage[latin1]{inputenc}
\usepackage{amssymb,amsbsy,amsmath}
\usepackage{bbm}
\newcommand{\op}[1]{\fontdimen12\textfont3=2pt\fontdimen12\scriptfont3=1.4pt\!\null\mathop{\protect\vphantom{#1}\smash{#1}}\limits_{\sim}\null\!}

\begin{document}



\title[CST]
      {Thermodynamics of the spin square}
\author{Heinz-J\"urgen Schmidt$^1$ and Christian Schr\"oder$^2$
}
\address{$^1$  Department of Physics,  Osnabr\"uck University,
 D - 49069 Osnabr\"uck, Germany\\
$^2$  Bielefeld Institute for Applied Materials Research, Bielefeld University of Applied Sciences, D - 33619 Bielefeld, Germany\\
and\\
Faculty of Physics, Bielefeld University, 33615 Bielefeld, Germany
}

\begin{abstract}
Small spin systems at the interface between analytical studies and experimental application have been intensively studied in recent decades.
The spin ring consisting of four spins with uniform antiferromagnetic Heisenberg interaction
is an example of a completely integrable system, in the double sense: quantum mechanical and classical.
However, this does not automatically imply that the thermodynamic quantities of the classical system can also be calculated explicitly.
In this work, we derive analytical expressions for the density of states, the partition function,
specific heat, entropy, and susceptibility. These theoretical results are confirmed by numerical tests.
This allows us to compare the quantum mechanical quantities for increasing spin quantum numbers $s$
with their classical counterparts in the classical limit $s\to \infty$.
As expected, a good agreement is obtained, except for the low temperature region.
However, this region shrinks with increasing $s$, so that the classical state variables emerge as envelopes of the quantum mechanical ones.

\end{abstract}

\maketitle

\section{Introduction}\label{sec:Intro}

Small spin systems have been studied theoretically for two main reasons.
In some cases, they can be realized by magnetic molecules and analyzed by measurements, e.g.,
of magnetization curves, which allows comparison with theoretical predictions \cite{Petal07,GOB14,QZYNSWZ17,BRT18,QZYNSZ21}.
Another reason is that some small spin systems can be described analytically
and thus can be used to test numerical calculations or theoretical conjectures,
either quantum-theoretically \cite{K97,K98,MSL99,MSSL00,BSS00,S13}
or classically \cite{LLB98,CLAL99,MSL99,MSSL00,C00,KL01,AK02,C07,SSHL15,S21,C22,SS22}.
In the case of the spin square, primarily the second motive applies,
since it seems to be very difficult to chemically synthesize molecules with perfect spin squares.

It is well known that the spin square with constant Heisenberg interaction and
arbitrary spin quantum number $s$ is analytically solvable due to the additional
conserved quantities of the partial spin sums corresponding to the diagonals.
The square belongs to a class of integrable spin systems where the Hamiltonian is a linear combination
of squared partial spin sums obtained by an iterative coupling scheme \cite{SS09}.
Other systems of this class are the bow-tie and the octahedron.
Nevertheless, the explicit calculation of thermodynamical quantities as, e.g.,
the specific heat or the susceptibility becomes more and more difficult for large $s$,
due to the increasing number of terms,
even if one uses computer-algebraic software. An obvious way out would be to consider the
classical limit $s\to \infty$, see \cite{L73,FKL07}.
Here one encounters the difficulty that,
while the time evolution of the classical spin square can be derived in an elementary way, see Section \ref{sec:GD},
there is, to our best knowledge, no explicit representation of the mentioned thermodynamic quantities in the literature.
The closest thing to this is a series representation for the partition function of a spin ring \cite{J67}
and an integral representation for the partition function of the square \cite{CLAL99}.
Following \cite{J67} simplified expressions of the partition function of the $N$-chain  have been
derived \cite{CGS08}, especially for $N=2,3,4$.

The paper is organized as follows. After recalling some general definitions and results in Section \ref{sec:GD}
we briefly explain, in Section \ref{sec:CNM}, the numerical methods used to evaluate classical thermodynamical functions in our case.
Then we calculate a closed analytical expression for the density of states (dos) function
of the classical spin square in Section \ref{sec:DS}.
This is accomplished by replacing the counting of quantum spin states by integrals in the limit $s\to \infty$.
Consequently,
the partition function and the specific heat can be obtained by integrations using computer-algebraic tools,
see Section \ref{sec:SH}, analogously for the entropy, see Section \ref{sec:E}.
The results are somewhat complicated, but can still be presented explicitly.
Similar integrations lead to the zero field static susceptibility, see Section \ref{sec:SU}.
All these expressions will be checked by comparison with numerical calculations using Monte Carlo and Wang-Landau techniques.
In addition, we will make comparisons between the classical quantities and the corresponding quantum analogues, for increasing $s$.
This will allow us to visualize the classical limit directly. We will summarize our findings in Section \ref{sec:SO}.

Comparing our results with the above mentioned series representation in \cite{J67},
we obtain a certain remarkable identity for a series with modified spherical Bessel functions.
This and related identities have been moved to the appendix \ref{sec:CO}.

\section{General definitions and results concerning classical time evolution}\label{sec:GD}

We consider a classical spin system described by four spin vectors ${\mathbf s}_\mu,\;\mu=1,\ldots,4,$ of unit length.
The total spin vector will be written as
\begin{equation}\label{deftotal}
{\mathbf S}:= {\mathbf s}_1+ {\mathbf s}_2+ {\mathbf s}_3+ {\mathbf s}_4
\;,
\end{equation}
and its length as $S$. Analogously, we define the partial sums of spins
\begin{eqnarray}\label{defpartial13}
{\mathbf S}_a&:=& {\mathbf s}_1+{\mathbf s}_3\;,\\
\label{defpartial24}
{\mathbf S}_b&:=& {\mathbf s}_2+{\mathbf s}_4
\;,
\end{eqnarray}
with lengths $S_a$ and $S_b$, resp.~, such that ${\mathbf S}={\mathbf S}_a+{\mathbf S}_b$.
The general Heisenberg Hamiltonian  will be written as
\begin{eqnarray}\label{defHam1}
 H&=&J\left( {\mathbf s}_1 \cdot  {\mathbf s}_2 +  {\mathbf s}_2 \cdot  {\mathbf s}_3 + {\mathbf s}_3 \cdot  {\mathbf s}_4+ {\mathbf s}_4 \cdot  {\mathbf s}_1\right)\\
 \label{defHam2}
 &=& \frac{J}{2}\left( S^2-S_a^2-S_b^2\right)
 \;,
\end{eqnarray}
with a positive coupling coefficient $J>0$ that will be set to $1$ in what follows.
The possible energies range from $E=-4$ to $E=4$,
such that the ground state with $E_{\text{min}}=-4$ will be the N\'{e}el state $\uparrow\downarrow\uparrow\downarrow$
and the maximal energy $E=4$ is assumed by the FM state $\uparrow\uparrow\uparrow\uparrow$.

More generally, the partial spin inversion $ {\mathbf s}_2\mapsto -{\mathbf s}_2,\;{\mathbf s}_4\mapsto -{\mathbf s}_4$
transforms a state of energy $E$ into a state of energy $-E$. This implies that the density of states to be calculated in the next section
will be an even function of $E$ that, according to the above remarks, vanishes outside the interval $[-4,4]$.

$H$ yields the corresponding Hamiltonian equations of motion, see \cite{S21},
\begin{eqnarray}\label{eom1}
  \dot{\mathbf s}_1&=& \left( {\mathbf s}_2+ {\mathbf s}_4\right)\times {\mathbf s}_1,\\
  \label{eom2}
  \dot{\mathbf s}_2&=& \left( {\mathbf s}_1+ {\mathbf s}_3\right)\times {\mathbf s}_2,\\
  \label{eom3}
  \dot{\mathbf s}_3&=& \left( {\mathbf s}_2+ {\mathbf s}_4\right)\times {\mathbf s}_3,\\
  \label{eom3}
  \dot{\mathbf s}_4&=& \left( {\mathbf s}_3+ {\mathbf s}_1\right)\times {\mathbf s}_4,\\
  \;.
\end{eqnarray}

These equations of motion admit the conserved quantities $S_a^2, S_b^2, S^2, M:={\mathbf S}^{(3)},$
which satisfy the inequalities
\begin{eqnarray}
\label{ineq1}
 -S &\le& M\le S\;, \\
 \label{ineq2}
 \left|S_a-S_b\right| &\le& S\le S_a+S_b
 \;.
\end{eqnarray}

\begin{figure}[htp]
\centering
\includegraphics[width=0.6\linewidth]{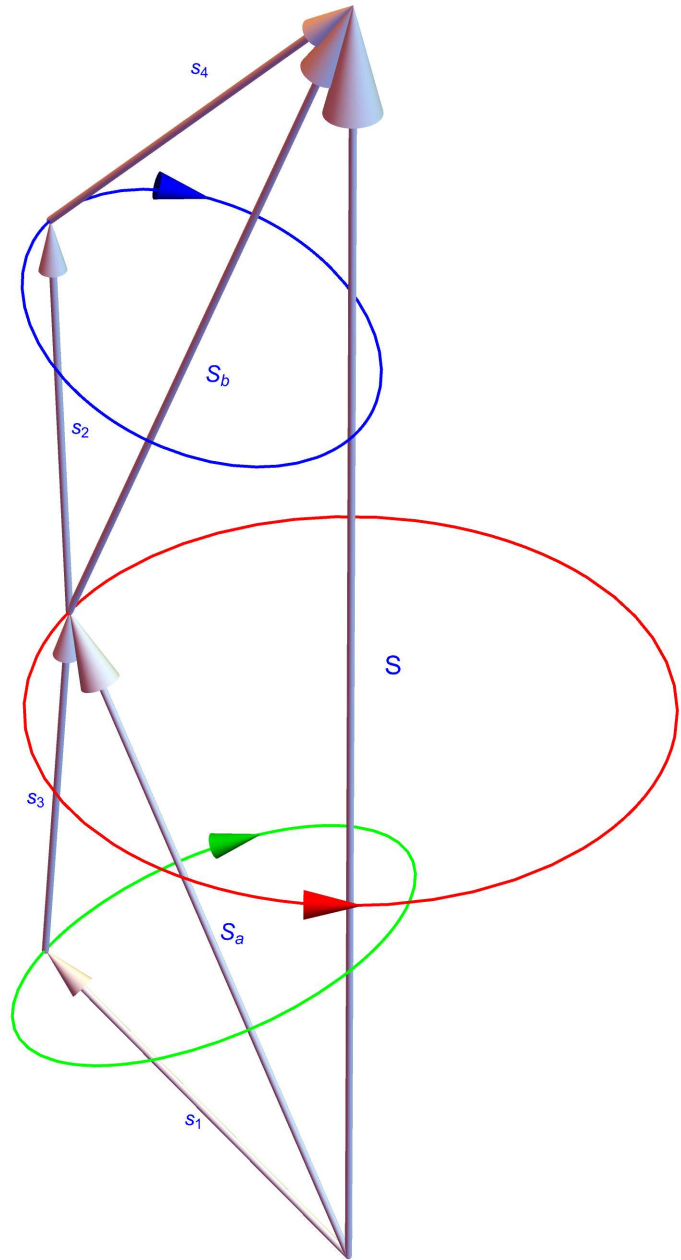}
\caption{Illustration of the time evolution of the spin square consisting of a superposition of different rotations.
The partial sum vectors ${\mathbf S}_a$ and  ${\mathbf S}_b$ rotate anti-clockwise around the total conserved spin vector
${\mathbf S}$ with angular velocity $S$ (red circle). The individual spin vectors ${\mathbf s}_1$ and  ${\mathbf s}_3$
in turn rotate clockwise around ${\mathbf S}_a$ with angular velocity $S_a$ (green circle), analogously ${\mathbf s}_2$ and  ${\mathbf s}_4$
around ${\mathbf S}_b$ (blue circle).
}
\label{FIGrot}
\end{figure}

Moreover, all three components of ${\mathbf S}$ are conserved as it holds for every Heisenberg spin system.
The partial spin sums satisfy the equations of motion
\begin{equation}\label{eqSa}
 \dot{\mathbf S}_a\stackrel{(\ref{eom1},\ref{eom3})}{=}
 \left( {\mathbf s}_2+ {\mathbf s}_4\right)\times {\mathbf s}_1+ \left( {\mathbf s}_2+ {\mathbf s}_4\right)\times {\mathbf s}_3
 = {\mathbf S}_b \times {\mathbf S}_a= {\mathbf S}\times {\mathbf S}_a
 \;,
\end{equation}
and, analogously,
\begin{equation}\label{eqSb}
 \dot{\mathbf S}_b= {\mathbf S}\times {\mathbf S}_b
 \;.
\end{equation}
This means that ${\mathbf S}_a$ and ${\mathbf S}_b$ will rotate about the fixed vector ${\mathbf S}$ with angular
velocity $S$. In a correspondingly rotating frame ${\mathbf s}_1$ and  ${\mathbf s}_3$ in turn will rotate about ${\mathbf S}_a$
with angular velocity $-S_a$. This follows from
\begin{equation}\label{eqs1}
 \dot{\mathbf s}_1\stackrel{(\ref{eom1})}{=} \left( {\mathbf s}_2+ {\mathbf s}_4\right)\times {\mathbf s}_1= {\mathbf S}_b\times {\mathbf s}_1
 =\left({\mathbf S}-{\mathbf S}_a \right)\times {\mathbf s}_1
 \;,
\end{equation}
and, analogously,
\begin{equation}\label{eqs1}
 \dot{\mathbf s}_3 =\left({\mathbf S}-{\mathbf S}_a \right)\times {\mathbf s}_3
 \;.
\end{equation}
In the same manner,  ${\mathbf s}_2$ and  ${\mathbf s}_4$  will rotate about the rotating ${\mathbf S}_b$
with angular velocity $-S_b$, see Figure \ref{FIGrot}.

\section{Classical numerical methods}\label{sec:CNM}
For a numerical investigation of the classical properties of the AF spin square we have performed standard
Monte-Carlo spin dynamics simulations \cite{ES12}.
One obtains the specific heat from such a simulation by calculating the fluctuations of the energy equation (\ref{defHam1}) according to
\begin{equation}\label{num_c}
	c(\beta)=\beta^2 \left(\left< E^2 \right> - \left<E \right>^2 \right)
\;,
\end{equation}
cp.~(\ref{defc}).
Here, $\left<...\right>$ denotes sample averaging during a Monte-Carlo simulation at a given {inverse} temperature $\beta$.
Likewise, one obtains the susceptibility by calculating the fluctuations of the total magnetization ${\mathbf S}$ according to
\begin{equation}\label{num_chi}
	\chi(\beta)=\frac{\beta}{3} \left(\left< {\mathbf S}^2 \right> - \left<{\mathbf S} \right>^2 \right)
\;,
\end{equation}
cp.~(\ref{chi2}).
For the numerical calculation of the classical density-of-states (dos) function $D(E)$ we have used the method developed by Wang and Landau \cite{WL01}. One advantage of this method is that once the density of states is known one can directly calculate thermodynamic quantities for any temperature without the need of performing any further simulations. Hence, by numerical integration of the dos in equation (\ref{defZ}) we obtain the temperature-dependent partition function $Z(\beta)$ as well as the inner energy $U_1(\beta) = \left<E \right>$ and the 2nd moment $U_2(\beta)=\left<E^2\right>$ by evaluating equation (\ref{defUn}). Given this, one can directly calculate the temperature-dependent specific heat using equation (\ref{num_c}) or equation (\ref{defc}), resp. and the entropy using equation (\ref{defS}).

\section{Density of states}\label{sec:DS}
It turns out that the classical density-of-states (dos) function $D(E)$ can be obtained in closed, even quite simple form.
To explain our method of calculating $D(E)$, it is appropriate to make a brief digression into the quantum theory of the AF spin square.

\subsection{Quantum case}\label{sec:DSQ}

The quantum version of the equation (\ref{defHam2}) can be written as
\begin{eqnarray}\label{defHamq}
 \op{H}=\frac{1}{2}\left( \op{\mathbf S}^2-\op{{\mathbf S}_a}^2-\op{{\mathbf S}_b}^2\right)
 \;,
\end{eqnarray}
where the Hermitean operators $\op{H},\op{\mathbf S}$ etc.~are defined on a $\left(2s+1\right)^4$-dimensional Hilbert space ${\mathcal H}$
and $s=1/2, 1, 3/2,\ldots$ is the individual spin quantum number. A convenient base of ${\mathcal H}$ is given by the product base, the vectors of which
being written as $\left| {\sf m}_1,{\sf m}_2,{\sf m}_3,{\sf m}_4\right\rangle$ where ${\sf m}_i=-s,-s+1,\ldots, s-1,s$  for $i=1,\ldots,4$.
The ${\sf m}_i$ are the eigenvalues of individual spin operators $\op{\mathbf S}_i^{(3)}$.

The coupling of the two spins of number $1$ and $3$ gives rise to another basis in the $\left(2s+1\right)^2$-dimensional Hilbert space ${\mathcal H}_{13}$
denoted by
\begin{equation}\label{CL13}
 \left| {\sf S}_a, {\sf m}_a \right\rangle = \sum_{{\sf m}_1,{\sf m}_3}
 CG\left( {\sf S}_a, {\sf m}_a ;s, {\sf m}_1,s,{\sf m}_3 \right) \left| {\sf m}_1,{\sf m}_3\right\rangle
 \;,
\end{equation}
where the $CG(\ldots)$ are the Clebsch-Gordan coefficients, see, e.~g., \cite{AS72}, 27.~9.~.
This is the common eigenbasis of $\op{\mathbf S}_a^2$ and $\op{\mathbf S}_a^{(3)}$ with  eigenvalues
$ {\sf S}_a\left({\sf S}_a+1 \right)$ and ${\sf m}_a$, resp.~. Analogous remarks apply to
the coupling of the two spins of number $2$ and $4$ which yields
\begin{equation}\label{CL24}
 \left| {\sf S}_b, {\sf m}_b \right\rangle = \sum_{{\sf m}_2,{\sf m}_4}
 CG\left( {\sf S}_b, {\sf m}_b ;s, {\sf m}_2,s,{\sf m}_4 \right) \left| {\sf m}_2,{\sf m}_4\right\rangle
 \;.
\end{equation}
Finally, we consider the coupling $(1,3),(2,4) \mapsto (1,2,3,4)$ which yields a common eigenbasis of
$\op{\mathbf S}^2, \op{\mathbf S}_a^2,\op{\mathbf S}_b^2,\op{\mathbf S}^{(3)}$
with eigenvalues $ {\sf S}\left(  {\sf S}+1\right),{\sf S}_a\left(  {\sf S}_a+1\right),{\sf S}_b\left(  {\sf S}_b+1\right),{\sf M}$, resp.~:
\begin{eqnarray}
\label{CL1234a}
 \left| {\sf S},  {\sf S}_a, {\sf S}_b,{\sf M}\right\rangle&=&
 \sum_{{\sf m}_a,{\sf m}_b}
 CG\left( {\sf S}, {\sf M} ; {\sf S}_a, {\sf m}_a, {\sf S}_b,{\sf m}_b \right) \left| {\sf S}_a, {\sf m}_a, {\sf S}_b,{\sf m}_b \right\rangle
\;.
\end{eqnarray}
These states are also eigenstates of the Hamiltonian (\ref{defHamq}) with eigenvalues
\begin{equation}\label{eigval}
 \varepsilon=\frac{1}{2}\left({\sf S}\left(  {\sf S}+1\right)-{\sf S}_a\left(  {\sf S}_a+1\right)-{\sf S}_b\left(  {\sf S}_b+1\right) \right)
 \;.
\end{equation}
Solving this equation for ${\sf S}\ge 0$ yields
\begin{equation}\label{SEq}
  {\sf S}={\sf S}_\varepsilon:=\frac{1}{2} \left(\sqrt{8 \varepsilon+(2 {\sf S}_a+1)^2+(2 {\sf S}_b+1)^2-1}-1\right)
  \;.
\end{equation}

This result can be used to calculate the number ${\mathcal N}_q({\sf E})$ of eigenstates with an energy $\varepsilon\le {\sf E}$.
Note that the Clebsch-Gordan coefficient (\ref{CL24}) vanishes if the triangle inequality
$\left| {\sf S}_a-{\sf S}_b\right|\le {\sf S}\le {\sf S}_a+{\sf S}_b$ is violated. For each ${\sf S}$ satisfying
this inequality and, additionally, ${\sf S}\le {\sf S}_{\sf E}$ there are $2{\sf S}+1$ states
with the same energy corresponding to different ${\sf M}$ in (\ref{CL1234a}). Hence
\begin{equation}\label{Nqe}
{\mathcal N}_q({\sf E}) =  \sum_{{0\le {\sf S}_a\le 2s}\atop{0\le {\sf S}_b\le 2s}}
\sum_{{\sf S}=\left| {\sf S}_a-{\sf S}_b\right|}^{\text{Min}\left( {\sf S}_{\sf E}, {\sf S}_a+{\sf S}_b\right)}\; 2{\sf S}+1
\;.
\end{equation}

\subsection{Classical case}\label{sec:DSC}

\begin{figure}[htp]
\centering
\includegraphics[width=0.7\linewidth]{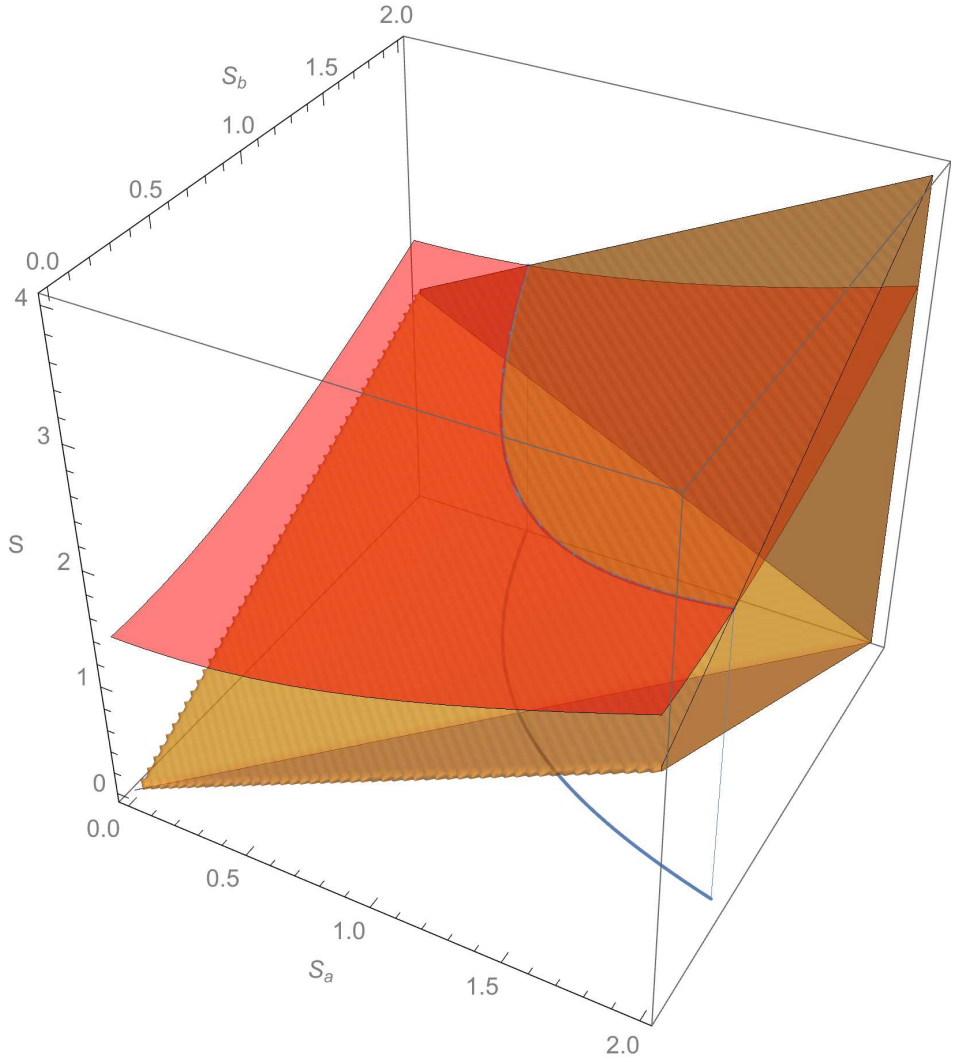}
\caption{The region in $(S_a,S_b,S)$-space for the integration (\ref{Nce}).
The dark yellow polyhedron represents the triangle inequality (\ref{ineq2}) and the light red surface is the graph of $S=S_E(S_a,S_b)$
for the example $E=1$. This graph intersects the boundary of the polyhedron at the blue curve which is projected onto
the $S_a-S_b$-plane and yields the hyperbola $S_a\,S_b =E$ according to (\ref{HypE}). The domain of integration (\ref{Nce})
will hence be the intersection of the polyhedron with the region below the surface $S=S_E$.
}
\label{FIGfr}
\end{figure}

\begin{figure}[htp]
\centering
\includegraphics[width=0.7\linewidth]{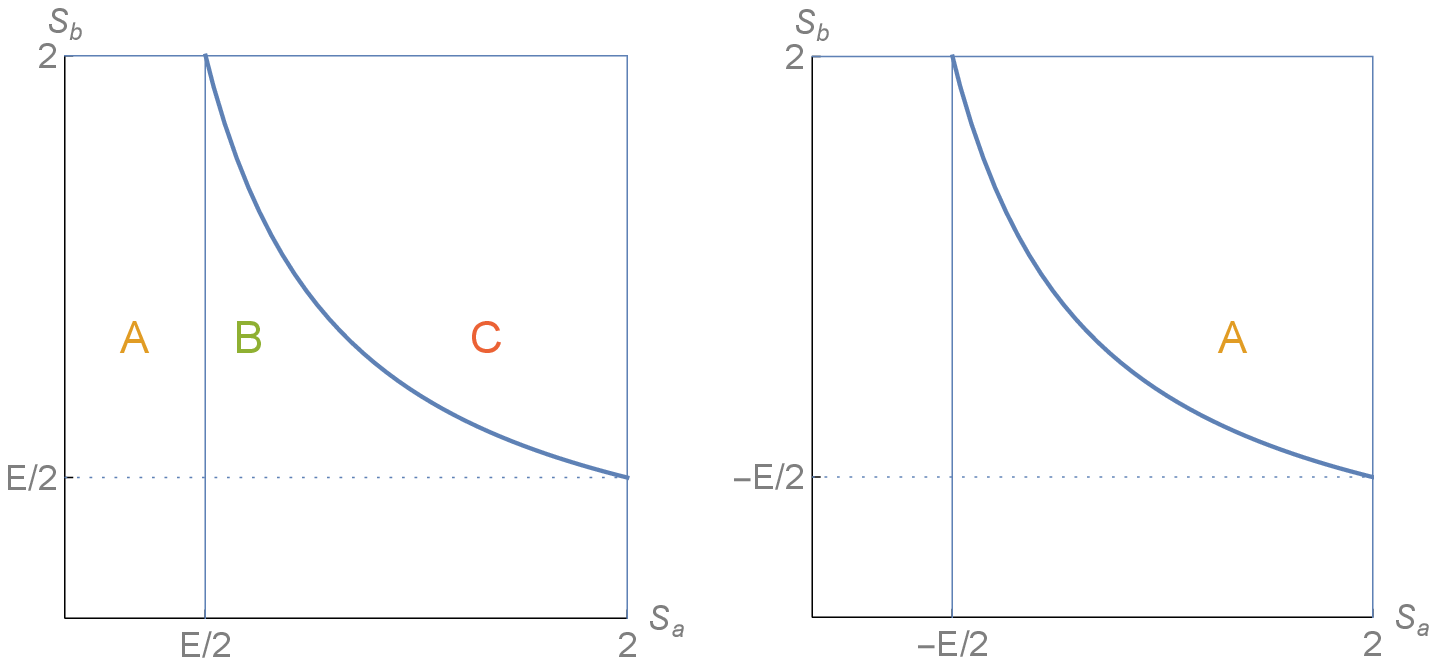}
\caption{The region in the $(S_a,S_b)$-plane for the integration (\ref{Nce}) divided by the hyperbola $S_a\,S_b =| E|$.
Left panel: $E>0$.
The integral (\ref{Nce}) assumes different forms in the three regions $A,B$ and $C$.
Right panel: $E<0$. The integral (\ref{Nce}) is only non-zero in the region $A$.
}
\label{FIGsq}
\end{figure}

Now we consider the classical limit $s\to\infty$.  This means: All quantum numbers are scaled according to
$S={\sf S}/s,\,S_a={\sf S}_a/s, E={\sf E}/(s(s+1)),\ldots$ etc.,
all functions are expanded into $1/s$-series and only the leading terms independent of $s$ are kept.
The function ${\sf S}_\varepsilon$ according to (\ref{SEq}) will hence be replaced by
\begin{equation}\label{SEc}
 S_E:=\sqrt{2 E+S_a^2+S_b^2}
  \;.
\end{equation}
The sums occurring in (\ref{Nqe}) are replaced by corresponding integrals resulting in the following expression
\begin{equation}\label{Nce}
{\mathcal N}(E) =  \int_{0}^{2}dS_a\,\int_{0}^{2}dS_b\,\int_{\left| S_a-S_b\right|}^{\text{Min}\left( S_E,S_a+S_b \right)}2\,S\,dS
\end{equation}
for the volume formed of phase space points with energy $\le E$.

Although the dos is symmetric w.~r.~t.~$E$ we will consider both cases $E>0$ and $E<0$ separately
in order to simplify the presentation of the analogous calculations for the susceptibility, see Section \ref{sec:SUC}.

\subsubsection{$E>0:$}

To simplify the evaluation of the minimum in the upper integral limit in (\ref{Nce}) we will divide the
$S_a-S_b$-square into two regions according to whether $S_a+S_b\le S_E$ or not. Note that
\begin{equation}\label{HypE}
 S_a+S_b\le S_E \Leftrightarrow \left(S_a+S_b \right)^2\le S_E^2=2E+S_a^2+S_b^2 \Leftrightarrow S_a\,S_b \le E
 \;.
\end{equation}
Hence the $S_a-S_b$-square is divided into two regions separated by a hyperbola, see Figure \ref{FIGfr}.

Next, we will perform the integration (\ref{Nce}) step by step.
The integral over $S$ has two different forms, depending
on whether  $S_a+S_b\le S_E$ or not. In the first case we obtain
\begin{equation}\label{int1}
I_1= \int_{\left|S_a-S_b \right|}^{S_a+S_b} S\,dS = 2 S_a\, S_b
\;,
\end{equation}
where we have omitted the overall factor $2$ that can be absorbed by the normalization constant.
The second case yields
\begin{equation}\label{int2}
I_2= \int_{\left|S_a-S_b \right|}^{S_E} S\,dS = E+ S_a\, S_b
\;.
\end{equation}
The following integrations over $S_a$ and $S_b$ can be split into three parts corresponding to the regions $A,B$ and $C$, see Figure \ref{FIGsq}, left panel.
For the contribution from $A$ we obtain
\begin{equation}\label{IA}
 I_A\stackrel{(\ref{int1})}{=} \int_{0}^{E/2}dS_a\left(\int_0^2 dS_b (2 \,S_a\, S_b) \right)=\frac{E^2}{2}
 \;.
\end{equation}
The region $B$ contributes the integral
\begin{equation}\label{IB}
 I_B\stackrel{(\ref{int1})}{=} \int_{E/2}^{2}dS_a\left(\int_0^{E/S_a} dS_b (2 \,S_a\, S_b) \right)=E^2 \log \left(\frac{4}{E}\right)
 \;.
\end{equation}
In the region $C$ the upper bound of $S$ will be $S_E$ and hence the corresponding integral reads
\begin{equation}\label{IC}
 I_C\stackrel{(\ref{int2})}{=} \int_{E/2}^{2}dS_a\left(\int_{E/S_a}^{2} dS_b (E+S_a\, S_b) \right)=
 4+4 E-\frac{5 E^2}{4}-\frac{3}{2} E^2 \log \left(\frac{4}{E}\right)
 \;.
\end{equation}
The sum of (\ref{IA}),  (\ref{IB}) and  (\ref{IC}) can be simplified to
\begin{equation}\label{IABC}
 {\mathcal N}(E)=I_A+I_B+I_C=C\left(-\frac{3 E^2}{4}-\frac{1}{2} E^2 \log \left(\frac{4}{E}\right)+4 E+4\right)
 \;,
\end{equation}
where the normalization constant $C$ is preliminarily left open.
The derivative of ${\mathcal N}(E)$ gives the dos:
\begin{equation}\label{dos1}
 D(E)=\frac{\partial  {\mathcal N}(E)}{\partial E}=C\left(4-E \left(1+\log \left(\frac{4}{E}\right)\right)\right),\mbox{ for } 0\le E\le 4
 \;.
\end{equation}

\subsubsection{$E<0:$}
For $E<0$ we have
\begin{equation}\label{inequSE1}
 S_E=\sqrt{2E+S_a^2+S_b^2}<\sqrt{S_a^2+S_b^2}\le S_a+S_b
 \;,
\end{equation}
and hence the upper integral limit in (\ref{Nce}) will always be $S_E$.
However, it may happen that the lower integral limit in (\ref{Nce}) is above $S_E$,
in which case the integral (\ref{Nce}) vanishes. This occurs iff
\begin{equation}\label{inequSE2}
S_E \le \left|S_a-S_b\right| \Leftrightarrow 2E+S_a^2+S_b^2 \le S_a^2-2 S_a S_b+S_b^2 \Leftrightarrow S_a S_b \le -E = \left| E\right|
\;,
\end{equation}
that is, iff $(S_a,S_b)$ lies outside the region $A$, see Figure \ref{FIGsq}, right panel.
Therefore the only contribution to (\ref{Nce}) will be given by
\begin{equation}\label{IIA}
 {\mathcal N}(E)\stackrel{(\ref{int2})}{=} \int_{-E/2}^{2}dS_a\left(\int_{-E/S_a}^{2} dS_b (E+S_a\, S_b) \right)=
\frac{3 E^2}{4}-\frac{1}{2} E^2 \log \left(-\frac{E}{4}\right)+4 E+4
 \;.
\end{equation}
The derivative of ${\mathcal N}(E)$ gives the dos:
\begin{equation}\label{dos2}
 D(E)=\frac{\partial  {\mathcal N}(E)}{\partial E}=C\left(4+E \left(1+\log \left(-\frac{E}{4}\right)\right)\right),\mbox{ for } -4\le E\le 0
 \;.
\end{equation}
This agrees with (\ref{dos1}) if $E$ is replaced by $-E$, thereby again confirming the symmetry of the dos.

\subsubsection{$-4\le E\le 4:$}

\begin{figure}[htp]
\centering
\includegraphics[width=0.7\linewidth]{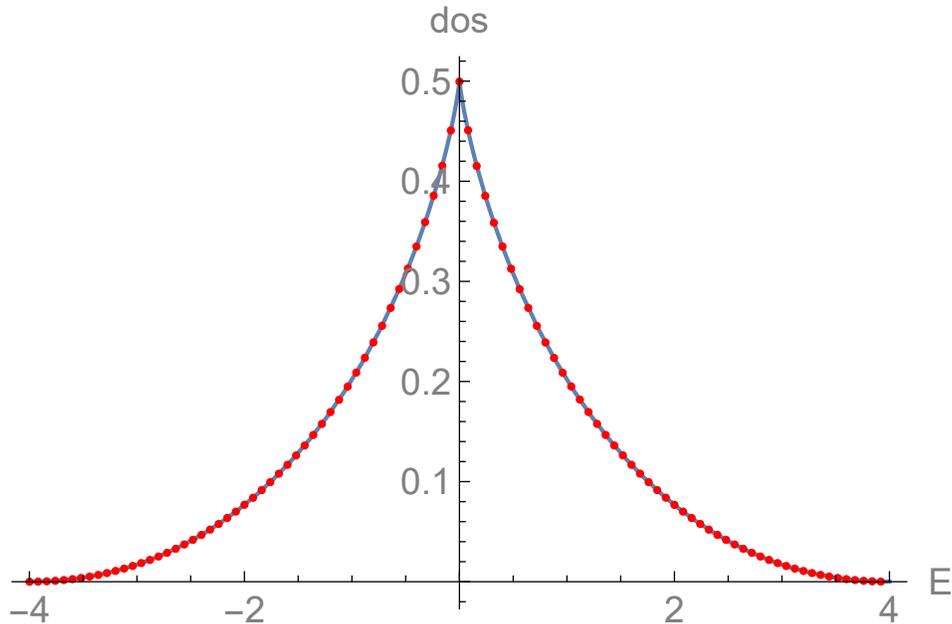}
\caption{Comparison of the analytical density of states (dos) function
$D(E)$ according to (\ref{dos3}) (blue curve) with the numerical result (red points) obtained by Wang-Landau sampling \cite{WL01}.
}
\label{FIGld}
\end{figure}

Extending $D(E)$ to all values of $E$ and calculating the normalization constant as $C=1/8$ finally gives the result
\begin{equation}\label{dos3}
 D(E)=  \left\{\begin{array}{l@{\quad:\quad}l}
 \frac{1}{8} \left(4-| E|  \left(1+\log \left(\frac{4}{| E| }\right)\right)\right) &  -4\le E\le 4,
 \\
0 & \mbox{else},
 \end{array} \right.
\end{equation}
see Figure \ref{FIGld}.

For later purposes we expand the dos at $E=E_{\text{min}}=-4$ into a power series in the variable $\epsilon=E-E_{\text{min}}=E+4$:
\begin{eqnarray}\label{doser1}
D(E)&=&D(-4+\epsilon)=\sum_{n=2}^{\infty}\frac{2^{-2 n-1}}{(n-1) n}\,\epsilon^n\\
&=&\frac{\epsilon ^2}{64}+\frac{\epsilon ^3}{768}+\frac{\epsilon ^4}{6144}+\frac{\epsilon
   ^5}{40960}+\frac{\epsilon ^6}{245760}+\frac{\epsilon ^7}{1376256}+\frac{\epsilon
   ^8}{7340032}+\frac{\epsilon ^9}{37748736}+\frac{\epsilon ^{10}}{188743680}+O\left(\epsilon
   ^{11}\right).
\end{eqnarray}

\section{Specific heat}\label{sec:SH}

\subsection{Classical case}\label{sec:SHC}

For the calculation of the specific heat we do not have to resort to the phase space, but can
use the density-of-states function $D(\varepsilon)$ modulated by the Boltzmann factor $e^{-\beta\,\varepsilon}$,
where $\beta=1/T$ denotes the (dimensionless) inverse temperature, as usual.
Then the partition function can be expressed as
\begin{equation}\label{defZ}
 Z(\beta)= \int_{-4}^{4} D(\varepsilon)\,e^{-\beta\,\varepsilon}\,d\varepsilon
 \;.
\end{equation}
Using the explicit form of $D(\varepsilon)$ according to (\ref{dos3}) we obtain
\begin{equation}\label{Zanal}
 Z(\beta)=\frac{1}{4 \beta ^2}\left(\text{Chi}(4 \beta )-\log (4 \beta )-\gamma \right)
 \;,
\end{equation}
where $\text{Chi}(z)$ denotes the hyperbolic cosine integral, see \cite{NIST21}, 6.~2.~16.~, and $\gamma$ the Euler constant.

The inner energy is obtained as the first moment $U_1(\beta)$ where the $n-$th moment is defined by
\begin{equation}\label{defUn}
U_n(\beta)=\frac{1}{Z(\beta)}\,\int_{-4}^{4}\varepsilon^n\, D(\varepsilon)\,e^{-\beta\,\varepsilon}\,d\varepsilon
\;.
\end{equation}
For the spin square the inner energy assumes the form
\begin{equation}\label{U1}
 U_1(\beta)=\frac{1}{\beta }\left(2+\frac{2 \sinh ^2(2 \beta )}{\log (\beta )-\text{Chi}(4 \beta )+\gamma +\log(2)}\right)
 \;.
\end{equation}

After a short calculation the specific heat $c(T):=\frac{\partial}{\partial T} U_1(1/T)$ can be written in the well-known form as
\begin{equation}\label{defc}
  c(\beta)=\beta^2 \left( U_2(\beta)-U_1^2(\beta)\right)
  \;,
\end{equation}
where $U_2(\beta)$ denotes the second moment according to the general definition (\ref{defUn}).
The explicit form reads
\begin{eqnarray}\nonumber
c(\beta)&=&\frac{1}{(\log (\beta )-\text{Chi}(4 \beta )+\gamma +\log (4))^2}\times\\
\nonumber
&&\left((\log (\beta )-\text{Chi}(4 \beta )+\gamma +\log (4)) (6 \log (\beta )
-4 \beta  \sinh (4 \beta)+5 \cosh (4 \beta )-6 \text{Chi}(4 \beta )+6 \gamma -5+12\log (2))\right.\\
\label{canal}
&&\left.-(2 \log (\beta )+\cosh (4\beta )-2 \text{Chi}(4 \beta )+2 \gamma -1+\log (16))^2\right)
\;.
\end{eqnarray}
Using the series expansion (\ref{doser1}) of the dos we can easily derive a low temperature expansion of $c(T)$ of the form
\begin{equation}\label{serc}
 c(T)\sim 3+\frac{T}{2}+\frac{9 T^2}{16}+\frac{13 T^3}{16}+\frac{355 T^4}{256}+\frac{1383
   T^5}{512}+\frac{24129 T^6}{4096}+\frac{29093 T^7}{2048}+\frac{2460087 T^8}{65536}-\frac{2182975
   T^9}{131072}-\frac{7263267 T^{10}}{1048576}+O\left(T^{11}\right)
   \;.
\end{equation}

The low temperature limit $\lim_{T\to 0}c(T)=\alpha+1=3$ is due to the lowest non-vanishing power of $\alpha=2$ in the expansion (\ref{doser1}).
It can also be derived by the following elementary argument: The number of degrees of freedom of the spin square is $2\times 4=8$
and the N\'{e}el ground state is collinear and can be freely rotated in spin space. This reduces the number of quadratic modes
for excitations close to the ground state to $8-2=6$. Each quadratic mode contributes $\frac{1}{2}$ to the specific heat at $T=0$
which confirms the above result of $\lim_{T\to 0}c(T)=3$.

The high temperature limit of $c(T)$ is obtained by a Taylor expansion of (\ref{canal}) in the variable $\beta=1/T$:
\begin{equation}\label{serchigh}
 c(T)\sim\frac{4}{3}\frac{1}{T^2}+\frac{8}{45} \frac{1}{T^4}-\frac{16}{63}
   \frac{1}{T^6}+\frac{6304}{91125} \frac{1}{T^8}+\frac{64}{66825}
   \frac{1}{T^{10}}+O\left(\frac{1}{T^{12}}\right)
   \quad\mbox{for } T\to \infty
   \;.
\end{equation}

\begin{figure}[htp]
\centering
\includegraphics[width=0.7\linewidth]{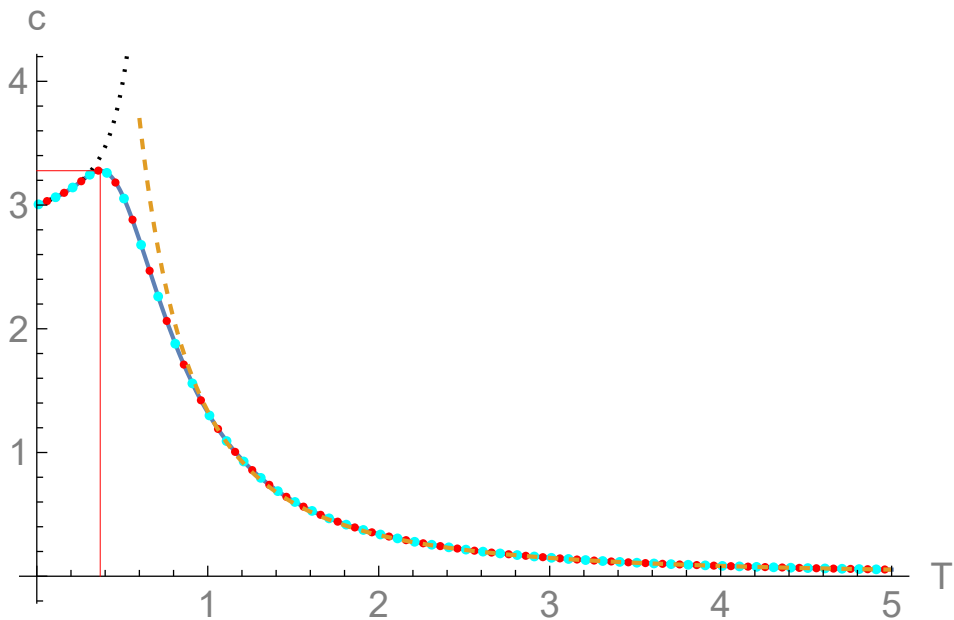}
\caption{Specific heat $c(T)$ as a function of temperature $T$. We show the analytical result (\ref{canal}) (blue curve), the low-temperature expansion
(\ref{serc}) (black, dotted curve), the leading power $\frac{4}{3 T^2}$ of the high temperature limit according to (\ref{serchigh}) (dashed, yellow curve)
as well as the numerical results using Monte-Carlo simulation (red points) and Wang-Landau calculation (cyan points).
Note the low temperature limit $c(T=0)=3$ as well as the global maximum at  $T_m=0.371046$ with height $c_m=c(T_m)=3.27751$ (thin red lines).
}
\label{FIGct}
\end{figure}

Hence $c(T)$ decays for $T\to\infty$ with the leading power of $\frac{4}{3 T^2}$.
Moreover, it has a global maximum at $T_m=0.371046$ with height $c_m=c(T_m)=3.27751$, see Figure \ref{FIGct}.

\subsection{Quantum case}\label{sec:SHQ}

Using the results of Section \ref{sec:DSQ} it is a straightforward task to calculate the eigenvalues $\varepsilon_n=\varepsilon_{S_a,S_b,S,M}$,
see (\ref{eigval}), together with their multiplicity $d_n$.
In general, further degeneracies occur, additional to the degeneracy w.~r.~t.~$M$,  which reduce the number of different eigenvalues.
For the largest chosen value of the individual quantum number $s=30$
we thus obtain $6103$ different eigenvalues with multiplicities $d_n$ that sum up to the dimension of the Hilbert space of $(2s+1)^4=13,845,841.$
Therefore, in this case, the partition function $Z=\sum_n d_n \exp( -\beta \varepsilon_n)$
has $6103$ terms and the specific heat $c(\beta)$ has about four times as many terms,
that can, nevertheless, be handled in reasonable computing times by a computer-algebraic software like MATHEMATICA.

\begin{figure}[htp]
\centering
\includegraphics[width=0.7\linewidth]{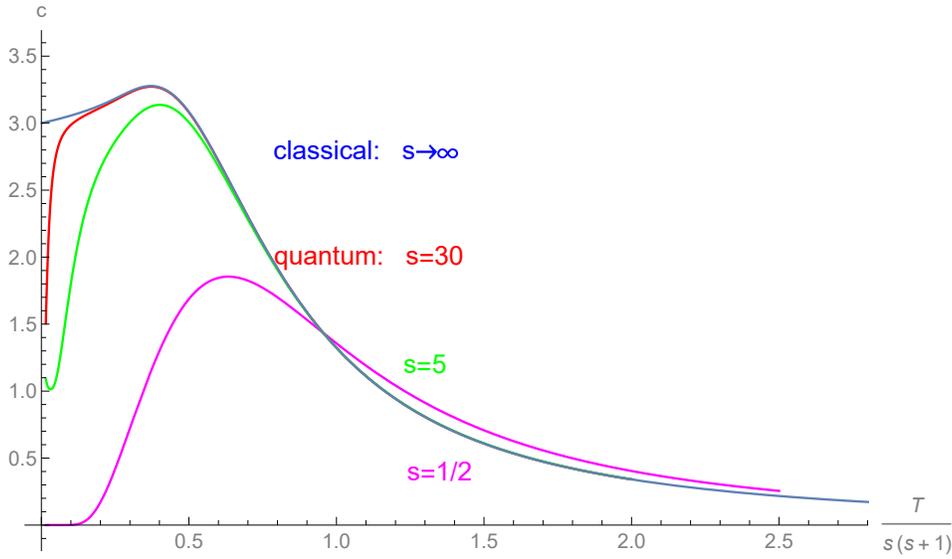}
\caption{Specific heat $c$ as a function of scaled temperature $\frac{T}{s(s+1)}$. We show the classical result (\ref{canal}) (blue curve),
as well as the quantum results for $s=1/2, 5,  30$ (magenta, green, red curves).
}
\label{FIGpcc}
\end{figure}

We will plot $c(T)$ for the quantum numbers $s=1/2, 5, 30$ as well as for the classical limit $s\to \infty$ calculated in the previous section.
For the sake of comparison we have chosen the scaled temperature $\frac{T}{s(s+1)}$ as the independent variable, see Figure \ref{FIGpcc}.
For a finite quantum spin system and  $T\to 0$, $c(T)$ always converges in a flat-foot fashion to $0$.
This can be clearly seen in Figure \ref{FIGpcc} for $s=1/2$ but not for $s=5$ or $s=30$ due to numerical instabilities.
However, Figure \ref{FIGpcc} also illustrates how this fact can be reconciled with $\lim_{T\to 0}c(T)=3$ in the classical case:
Besides the global maximum, the specific heat for the quantum systems at low temperatures and
increasing $s$ forms a shoulder and the classical $c(T)$ is the envelope of these shoulders for $s\to \infty$.

\section{Entropy}\label{sec:E}

\subsection{Classical case}\label{sec:EC}

The entropy $S(\beta)$ can be generally defined by
\begin{equation}\label{defS}
S(\beta):=\beta\,U_1(\beta)+\log Z(\beta)
\;.
\end{equation}
Inserting the above results for $U_1$, see (\ref{U1}), and $Z$, see (\ref{Zanal}),  we obtain
\begin{equation}\label{Sbeta}
  S(\beta)=2+\log \left(-\frac{\log (4 \beta )-\text{Chi}(4 \beta )+\gamma }{4 \beta ^2}\right)+\frac{2 \sinh
   ^2(2 \beta )}{\log (\beta )-\text{Chi}(4 \beta )+\gamma +\log (4)}
   \;.
\end{equation}
For $T\to 0$ the entropy diverges logarithmically. In order to determine the asymptotic form of $S(T)$ for $T\to 0$ we first consider
\begin{equation}\label{Chiasy}
  \text{Chi}(x) \sim \frac{1}{2}\exp(x)\left( \frac{1}{x} +\frac{1!}{x^2}  +\frac{2!}{x^3}+\frac{3!}{x^4}+\ldots \right) \quad \mbox{for } x\to \infty
  \;,
\end{equation}
which can be derived from \cite{NIST21}, 6.5.4., 6.2.20., 6.12.3., and 6.12.4.. Using $Z(\beta)\sim \frac{1}{4 \beta^2}\text{Chi}(4\beta)$ for $\beta\to \infty$
we thus obtain
\begin{equation}\label{logZasy}
  \log Z \sim \frac{4}{T} +3\log T - 5\,\log 2+\frac{T}{4} +\ldots\quad \mbox{for }T\to 0
  \;.
\end{equation}
Together with
\begin{equation}\label{betaUasy}
 \beta\, U_1 \sim -\frac{4}{T}+3+\frac{T}{4} +\ldots \quad \mbox{for }T\to 0
\end{equation}
we thus obtain for $T\to 0$
\begin{eqnarray}\nonumber
 S(T)&\sim& 3 \log T+(3-5 \log (2))+\frac{T}{2}+\frac{9 T^2}{32}+\frac{13 T^3}{48}\\
 \label{Sasy}
 && +\frac{355 T^4}{1024}+\frac{1383
   T^5}{2560}+\frac{8043 T^6}{8192}+\frac{29093 T^7}{14336}+\frac{2460087 T^8}{524288}-\frac{550015
   T^9}{1179648}+\frac{25395933 T^{10}}{10485760}+O\left(T^{11}\right)
   \;.
\end{eqnarray}

\begin{figure}[htp]
\centering
\includegraphics[width=0.7\linewidth]{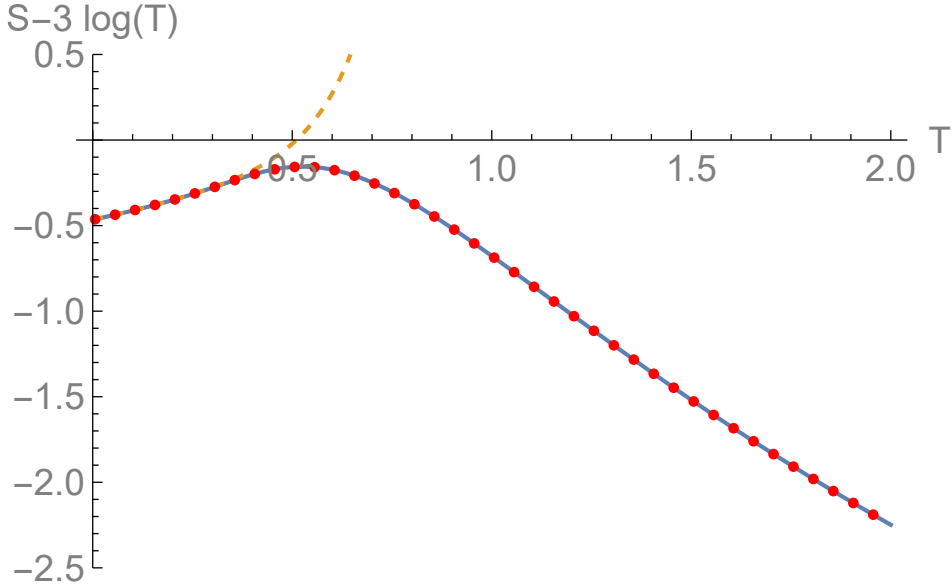}
\caption{Reduced classical entropy $S(T)-3\log T$ as a function of temperature $T$.
We show the analytical result (\ref{Sbeta}) (blue curve), the numerical Monte-Carlo calculations (red points) and the low-temperature expansion
(\ref{Sasy}) (yellow, dashed curve).
}
\label{FIGsl}
\end{figure}

Hence it is meaningful not to plot $S(T)$ but the {\em reduced entropy} $S(T)-3 \log T$, see Figure \ref{FIGsl}.

\subsection{Quantum case}\label{sec:EQ}

\begin{figure}[htp]
\centering
\includegraphics[width=0.7\linewidth]{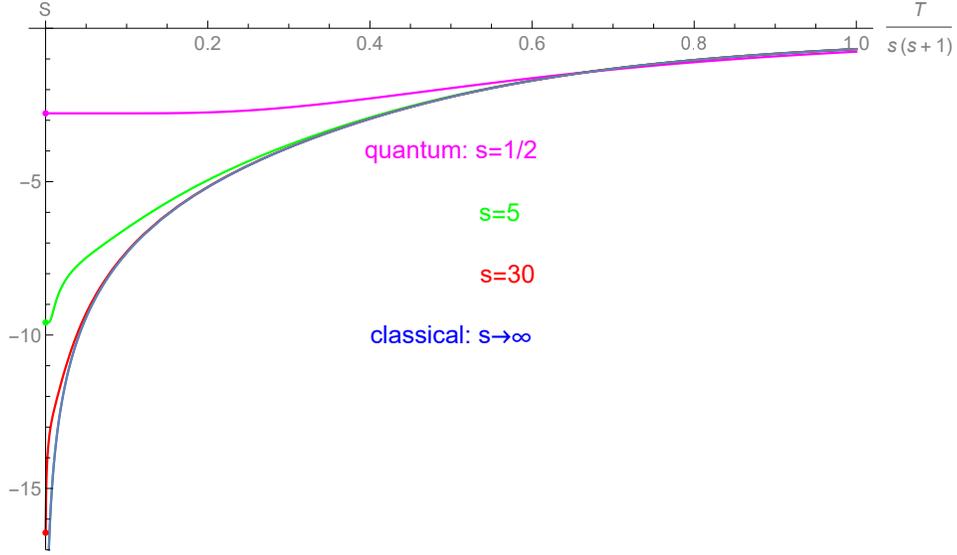}
\caption{Plot of the entropy $S$ as a function of scaled temperature $\frac{T}{s(s+1)}$.
We show the classical entropy $S_\text{cl}$ according to (\ref{Sbeta}) (blue curve) and the quantum entropies $S_\text{qu}$ for $s=1/2,5,30$
 (magenta, green, red curves), where the zero-point of $S_\text{qu}$ has been chosen as explained in the text.
The low temperature limits $S_\text{qu}(T\to 0)=-4\log (2s+1)$ are indicated by small colored dots.
}
\label{FIGsc}
\end{figure}

The calculation of the entropy $S_\text{qu}=\beta U_1 +\log(Z)$ in the quantum case is largely analogous to the calculation of the specific heat
explained in Section \ref{sec:SHQ}. However, there occurs an additional problem. In the high-temperature limit $\beta\to 0$ we have
\begin{equation}\label{limSbetato0}
 S_{\text{qu}}\to \lim_{\beta\to 0} \log Z = \lim_{\beta\to 0} \log \text{Tr} \exp(-\beta \op{H})= \log \text{Tr}\, {\mathbbm 1}= \log \text{dim}
 \;,
\end{equation}
where $\text{dim}=(2s+1)^4$ is the dimension of the Hilbert space.
This means that the high temperature limit depends on $s$. In contrast, the high temperature limit of the classical entropy
vanishes: $S_{\text{cl}}\to \lim_{\beta\to 0} \log Z =\log Z(0)= \log 1 =0$.
Therefore, approximate agreement of $ S_{\text{qu}}$ and $ S_{\text{cl}}$
for large temperatures can only be achieved by an $s$-dependent shift of the zero-point of entropy.
We decide to shift the quantum entropy and to replace $\log Z$  by $\log \frac{Z}{\text{dim}}$ in the definition of $S_\text{qu}$.
This has the consequence that the low temperature limit of $S_\text{qu}(T)$ no longer vanishes, as it would follow for a non-degenerate
ground state, but will be shifted to $S_\text{qu}(T\to 0)=-\log \text{dim}$. This is in accordance with the logarithmic divergence
of the classical entropy to $-\infty$, see Section \ref{sec:EC} and Figure \ref{FIGsc}.

\section{Susceptibility}\label{sec:SU}

\subsection{Classical case}\label{sec:SUC}

If we apply a magnetic field $B$ in $3$-direction, the Hamiltonian (\ref{defHam1}) is modified by an additional Zeeman term to
\begin{equation}\label{defHB}
 H_B= H - {\mathbf S}^{(3)} B
 \;.
\end{equation}
The resulting magnetization $M(B)$ is given by the expectation value of ${\mathbf S}^{(3)}$ using the modified  partition function.
Due to the isotropy of the Hamiltonian (\ref{defHam1}) the magnetization will vanish at $B=0$ and
hence the first generally non-vanishing term of the Taylor expansion of $M(B)$ at $B=0$ will be given by the
``zero field susceptibility"
\begin{equation}\label{defchi}
  \chi(\beta)=\left.\frac{d M}{dB}\right|_{B=0}
  \;.
\end{equation}
After a short calculation one obtains the well-known expression
\begin{equation}\label{chi1}
 \chi(\beta)= \frac{\beta}{Z_0}\int_{\mathcal P}\left({\mathbf S}^{(3)}\right)^2\,\exp\left( -\beta\,H\right)\,dV
 =:\beta\,\langle \left({\mathbf S}^{(3)}\right)^2\rangle
 \;.
\end{equation}
Again using isotropy of $H$ we have
$\langle \left({\mathbf S}^{(1)}\right)^2\rangle=\langle \left({\mathbf S}^{(2)}\right)^2\rangle=\langle \left({\mathbf S}^{(3)}\right)^2\rangle$
and thus
\begin{equation}\label{chi2}
 \chi(\beta) =\frac{\beta}{3}\langle S^2\rangle{=}\frac{\beta}{3 Z(\beta)} \int_{-4}^{4}\widetilde{D}(E)\exp(-\beta E) dE
 \;,
\end{equation}
where
\begin{equation}\label{chi3}
 \widetilde{D}(E)=\frac{\partial}{\partial E}\;\frac{1}{8}\,\int_{0}^{2}dS_a\,\int_{0}^{2}dS_b\,
 \int_{\left| S_a-S_b\right|}^{\text{Min}\left( S_E,S_a+S_b \right)}S^3\,dS
 \;,
\end{equation}
analogously to (\ref{Nce}). The factor $\frac{1}{8}$ has been introduced in (\ref{chi3}) according to the normalization constant
determined in (\ref{dos3}). Analogously to Section \ref{sec:DSC} we consider a case distinction according to the sign of $E$.

\subsubsection{$E>0:$}

To simplify the evaluation of the minimum in the upper integral limit in (\ref{chi3}) we will divide the
$S_a-S_b$-square into two regions according to whether $S_a+S_b\le S_E$ or not, separated by the hyperbola $S_a\,S_b=E$, see Figure \ref{FIGfr}.

The integral over $S$ has two different forms, depending
on whether  $S_a+S_b\le S_E$ or not. In the first case we obtain
\begin{equation}\label{sint1}
J_1= \int_{\left|S_a-S_b \right|}^{S_a+S_b} S^3\,dS =2 S_a S_b \left(S_a^2+S_b^2\right)
\;.
\end{equation}

The second case yields
\begin{equation}\label{sint2}
J_2= \int_{\left|S_a-S_b \right|}^{S_E} S^3\,dS =\left(S_a S_b+E\right) \left(-S_a S_b+S_a^2+S_b^2+E \right)
\;.
\end{equation}
The following integrations over $S_a$ and $S_b$ can be split into three parts corresponding to the regions $A,B$ and $C$, see Figure \ref{FIGsq}, left panel.
For the contribution from $A$ we obtain
\begin{equation}\label{JA}
 J_A\stackrel{(\ref{sint1})}{=} \int_{0}^{E/2}dS_a\left(\int_0^2 dS_b \,J_1 \right)=E^2+\frac{E^4}{16}
 \;.
\end{equation}
The region $B$ contributes the integral
\begin{equation}\label{JB}
 J_B\stackrel{(\ref{sint1})}{=} \int_{E/2}^{2}dS_a\left(\int_0^{E/S_a} dS_b \,J_1\right)=3 E^2-\frac{3 E^4}{16}
 \;.
\end{equation}
In the region $C$ the upper bound of $S$ will be $S_E$ and hence the corresponding integral reads
\begin{equation}\label{JC}
 J_C\stackrel{(\ref{sint2})}{=} \int_{E/2}^{2}dS_a\left(\int_{E/S_a}^{2} dS_b \,J_2 \right)=
 \frac{7 E^4}{48}+\frac{2}{3} E^3 \log (E)-\frac{4}{9} E^3 (2+\log (8))-2 E^2+\frac{32 E}{3}+\frac{80}{9}
 \;.
\end{equation}
The sum of (\ref{JA}),  (\ref{JB}) and  (\ref{JC}) can be simplified to
\begin{equation}\label{JABC}
 \widetilde{\mathcal N}(E)=J_A+J_B+J_C=
 \frac{E^4}{48}+\frac{2}{3} E^3 \log (E)-\frac{4}{9} E^3 (2+\log (8))+2 E^2+\frac{32
   e}{3}+\frac{80}{9}
 \;.
\end{equation}
The derivative of $\widetilde{\mathcal N}(E)$ gives the function
\begin{equation}\label{dos1}
 \widetilde{D}(E)=\frac{1}{8}\frac{\partial  \widetilde{\mathcal N}(E)}{\partial E}=
 \frac{1}{8}\left(\frac{E^3}{12}+2 E^2 \log (E)-E^2 (2+\log (16))+4 E+\frac{32}{3}\right),\mbox{ for } 0\le E\le 4
 \;.
\end{equation}

\subsubsection{$E<0:$}
 Analogously to Section \ref{sec:DSC} the only contribution to (\ref{chi3}) will be given by
\begin{equation}\label{JJA}
 \widetilde{\mathcal N}(E)\stackrel{(\ref{sint2})}{=} \int_{-E/2}^{2}dS_a\left(\int_{-E/S_a}^{2} dS_b \,J_2 \right)=
-\frac{E^4}{48}-\frac{2}{3} E^3 \log (-E)+\frac{4}{9} E^3 (2+\log (8))+6 E^2+\frac{32  E}{3}+\frac{80}{9}
 \;.
\end{equation}
The derivative of $\widetilde{\mathcal N}(E)$ gives the function
\begin{equation}\label{dos2}
\widetilde{D}(E)=\frac{1}{8}\frac{\partial  \widetilde{\mathcal N}(E)}{\partial E}=
-\frac{1}{8}\left(\frac{E^3}{12}-2 E^2 \log (-E)+E^2 (2+\log (16))+12 E+\frac{32}{3}\right)
,\mbox{ for } -4\le E\le 0
 \;.
\end{equation}

\subsubsection{$-4\le E\le 4:$}

\begin{figure}[htp]
\centering
\includegraphics[width=0.7\linewidth]{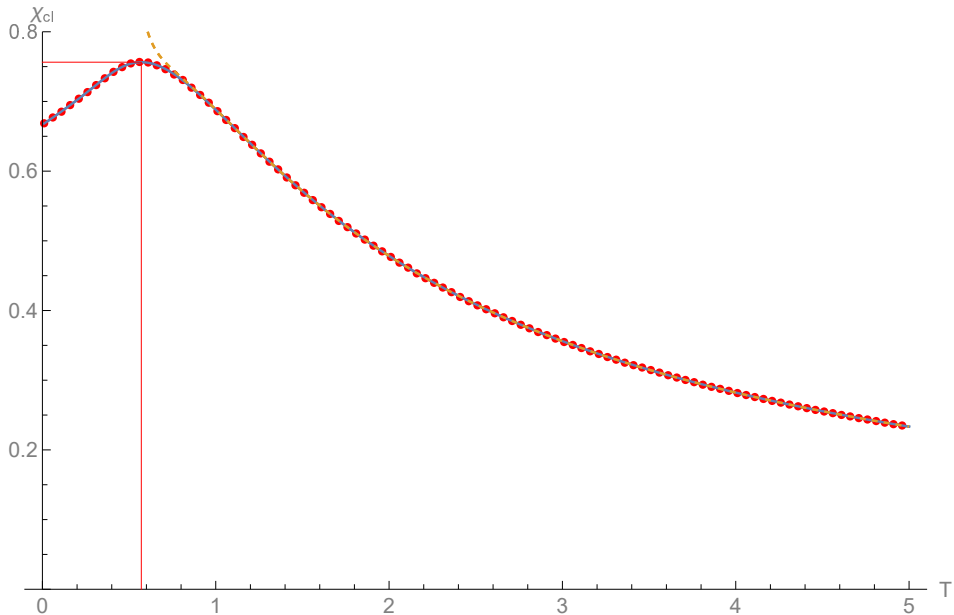}
\caption{Plot of the classical susceptibility $\chi_\text{cl}$ as a function of temperature $T$ according to (\ref{chianal}) (blue curve) together with the
numerical Monte-Carlo result (red points) and the
high temperature limit (\ref{chiser}) (dashed, yellow curve).  The limit of the classical susceptibility for vanishing $T$ is $\chi_\text{cl}(T\to 0)=2/3$.
$\chi_\text{cl}$ has a global maximum at $T_m=0.570563$ and $\chi_m=0.756339$ (thin, red lines).
}
\label{FIGchic}
\end{figure}

Extending $\widetilde{D}(E)$ to all values of $E$ finally gives the result
\begin{equation}\label{tildedos2}
\widetilde{D}(E)=  \left\{\begin{array}{l@{\quad:\quad}l}
\frac{1}{8}\left(-\frac{E^3}{12}-2 E^2 \log (-E)+E^2 (2+\log (16))+12 E+\frac{32}{3}\right)
&  -4\le E\le 0,
 \\
\frac{1}{8}\left(\frac{E^3}{12}+2 E^2 \log (E)-E^2 (2+\log (16))+4 E+\frac{32}{3}\right) &  0\le E\le 4,
 \\
0 & \mbox{else}.
 \end{array} \right.
\end{equation}
Inserting this result into (\ref{chi2}) and performing the integral gives, after some simplifications,
\begin{equation}\label{chianal}
 \chi(\beta)=\frac{4 \beta  (2 \beta +2 \gamma -1+\log (16))+8 \beta  \log (\beta )-4 \beta  \sinh (4
   \beta )+(4 \beta +1) \cosh (4 \beta )-8 \beta  \text{Chi}(4 \beta )-1}{6 \beta  (\log
   (\beta )-\text{Chi}(4 \beta )+\gamma +\log (4))}
\;,
\end{equation}
see Figure \ref{FIGchic}.
A Taylor expansion of (\ref{chianal}) at $\beta=0$ yields the high temperature limit
\begin{eqnarray}\nonumber
\chi(T)& \sim & \frac{4}{3 T}-\frac{8}{9} \frac{1}{T^2}
    +\frac{8}{27}  \frac{1}{T^3}-\frac{16}{405}\frac{1}{T^4}
    -\frac{16}{405}\frac{1}{T^5}+\frac{32}{945}\frac{1}{T^6}
   -\frac{416}{127575} \frac{1}{T^7}\\
   \label{chiser}
   && - \frac{12608}{1913625}\frac{1}{T^8}
   +\frac{1856 }{637875} \frac{1}{T^9}
   - \frac{128 }{1804275}\frac{1}{T^{10}}+O\left(\frac{1}{T^{11}}\right)
\;.
\end{eqnarray}

For $T\to 0$ the susceptibility is of the form $\frac{0}{0}$ since the total spin of the N\'{e}el ground state vanishes.
A closer inspection of (\ref{chianal}) shows that $\lim_{T\to 0}\chi(T)=2/3$, see Figure \ref{FIGchic}.

\begin{figure}[htp]
\centering
\includegraphics[width=0.7\linewidth]{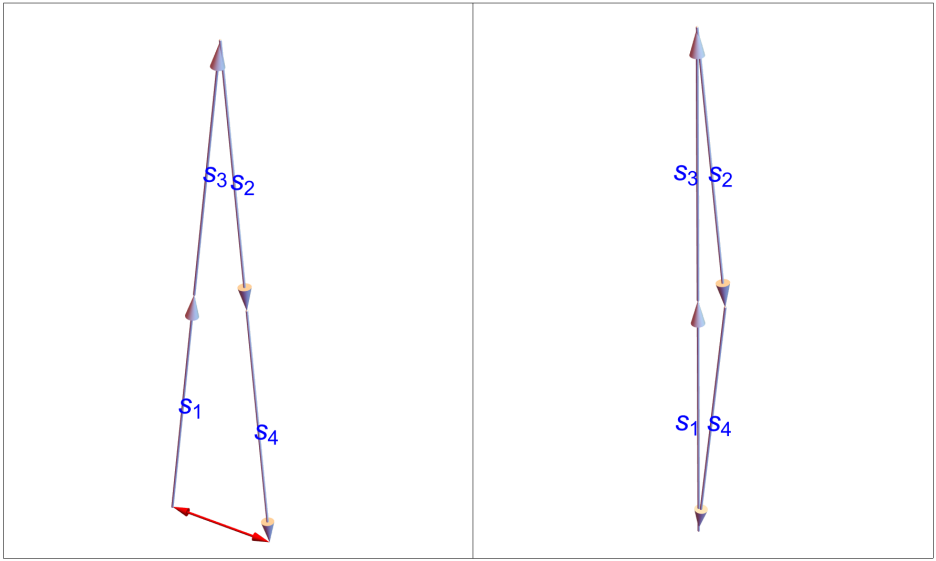}
\caption{Plot of two modes of excitation from the N\'{e}el ground state $\uparrow\downarrow\uparrow\downarrow$.
The first mode (left panel) leads to finite total spin square $S^2$ (red double arrow), whereas the total spin square $S^2$ of the second mode
(right panel) vanishes in quadratic order. There are two modes of the first kind (the one shown and its rotation by $90^\circ$)
and four modes of the second kind (resulting from interchanging the pairs (1,3) and (2,4) and rotations by $90^\circ$).
}
\label{FIGgg}
\end{figure}

It will be instructive to sketch another derivation of this result.
To this end we consider Cartesian coordinates $(x_\mu,y_\mu),\;\mu=1,2,3,4$ in the tangent planes at ${\mathbf s}_\mu$ for
a N\'{e}el ground state $\uparrow\downarrow\uparrow\downarrow$ and calculate the Hessian $K$ of the Hamiltonian $H$ w.~r.~t.~these coordinates.
$K$ has the eigenvalues $2,2,1,1,1,1,0,0$ corresponding to six quadratic modes and two zero modes describing rotations of the ground state.
The eigenvectors characterizing these modes give rise to an alternative set of ``normal" coordinates $\xi_i,\;i=1,\ldots,8$.
The total energy $H$ as well as the square of the total spin $S^2$ of a superposition of excitations can be expanded into a Taylor series in the $\xi_i$,
namely
\begin{eqnarray}
\label{ES2a}
  H &=& -4+ 2 \left( \xi_1^2+\xi_2^2\right)+ \sum_{j=3}^6\,\xi_j^2+O\left(\boldsymbol{\xi}^3\right),\\
  \label{ES2b}
  S^2 &=& 4 \left( \xi_1^2+\xi_2^2\right)+O\left(\boldsymbol{\xi}^3\right)
  \;.
\end{eqnarray}
This means that only the first two quadratic modes contribute to $S^2$ and hence to the susceptibility $\chi(T=0)$, see Figure \ref{FIGgg}.
We will calculate the contribution of the first mode:
\begin{eqnarray}\label{firstmode1}
 \left\langle S^2 \right\rangle &=& \int_{0}^{\infty}4 \xi_1^2\,\exp\left(-\beta(-4+2 \xi_1^2) \right)\,d\xi_1 \bigg/
 \int_{0}^{\infty}\exp\left(-\beta(-4+2 \xi_1^2) \right)\,d\xi_1\\
 &=& \frac{1}{\beta}
 \;.
\end{eqnarray}
In view of $\chi=\frac{\beta}{3}\left\langle S^2 \right\rangle$
the first mode contributes $1/3$ to the susceptibility; and analogously for the second mode which explains $\lim_{T\to 0}\chi(T)=2/3$.

\subsection{Quantum case}\label{sec:SUQ}
Also the calculation of the susceptibility $\chi_\text{qu}$ in the quantum case is largely analogous to the procedure for the specific heat
and for the entropy. The difference is that we have to calculate not only the different energy eigenvalues $\varepsilon_n$ and their degeneracies $d_n$
but have also to resolve for different total spin quantum numbers resulting in a number of triples $(d_m, \varepsilon_m,S_m)$.
For $s=20$ there are already $20,239$ such triples and corresponding numbers of terms for $\chi_\text{qu}$. Hence we restrict ourselves to
$s=20$ as the maximal spin quantum number.

\begin{figure}[htp]
\centering
\includegraphics[width=0.7\linewidth]{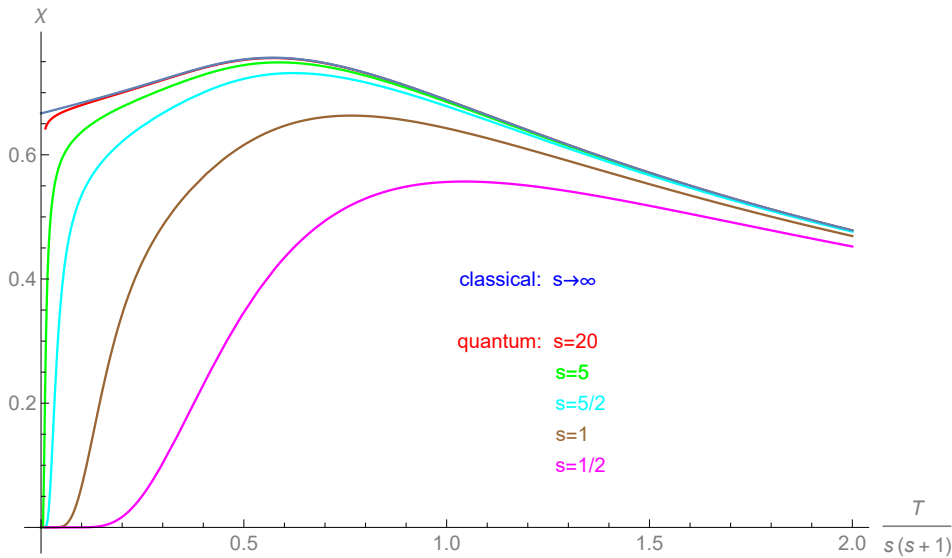}
\caption{Plot of the susceptibility $\chi$ as a function of scaled temperature $\frac{T}{s(s+1)}$.
We show the classical susceptibility $\chi_\text{cl}$ according to (\ref{chianal}) (blue curve)
and the quantum susceptibilities $\chi_\text{qu}$ for $s=1/2,1,5/2,5,20$
(magenta,  brown, cyan, green, red curves).
}
\label{FIGchi}
\end{figure}

Since the spin square has a unique ground state with $S=0$ the susceptibility $\chi_\text{qu}(T)$ converges to $0$, even in a flat-footed fashion.
This can be seen for $s=1/2,1,5/2,5$ in Figure \ref{FIGchi}. The classical susceptibility $\chi_\text{cl}(T)$ again appears as the envelope of
all $\chi_\text{qu}$, plotted versus scaled temperature $\frac{T}{s(s+1)}$, with the aforementioned finite limit $\chi_\text{qu}(T\to 0)=2/3$,
see Figure \ref{FIGchi}.

\section{Summary and Outlook}\label{sec:SO}

We have derived closed-form expressions for certain thermodynamic quantities of the classical AF spin square,
viz. density of states, partition function, specific heat, entropy, and susceptibility.
The partition function of the square has already been given in \cite{CLAL99}
albeit in terms of definite integrals, which could not be computed explicitly.
Our results were numerically verified by Monte Carlo simulations and/or Wang-Landau calculations.
The quantum version of the thermodynamic functions was also calculated analytically
using known identities for the coupling of four spins,
but these are too intricate to be stated explicitly.
For practical reasons, we have limited these calculations to quantum numbers of $s\le 20$ for susceptibility
and to $s\le 30$ for the other cases.

With analytical results for thermodynamic functions in both cases, quantum and classical,
a comparison with the classical limit $s\to \infty$ is obvious.
This comparison not only shows the consistency between the two classes of results,
but also illustrates the precise way to understand the classical limit.
The first point to pay attention to is the correct scaling of the temperature in the form $\frac{T}{s(s+1)}$,
which follows the scaling of the energy.
Another point is the behavior for $T\to 0$.
Quantum mechanically, the specific heat and susceptibility approach $0$ flatly,
while the classical counter-parts take positive values.
This apparent contradiction can be resolved by observing that the "quantum flat feet"
are increasingly compressed with respect to the scaled temperature,
yielding the classical quantities as their envelopes, see figures \ref{FIGpcc} and \ref{FIGchi}.

With respect to entropy, another apparent contradiction occurs: The quantum entropy at $T=0$ should vanish due to a non-degenerate
ground state, while the classical entropy diverges logarithmically to $-\infty$. As explained in detail in Section \ref{sec:E}
the classical limit of the quantum entropy for medium and large $T$ can only be reached if the high temperature limits
are properly adjusted. Here we have made the obvious choice of re-defining the quantum partition function by
$Z_\text{qu} = \text{Tr} \left(\exp(-\beta \op{H})/\text{dim}\right)$, which resolves the aforementioned contradiction in the low temperature region.

The comparison of our results with the series representation of the partition function in \cite{J67}
has led to certain identities, see the appendix \ref{sec:CO}.
For the spin triangle, it appears that the result of \cite{J67} needs to be modified,
a discovery that requires further investigation.
Another future task would be to extend the present results
to other similarly integrable spin systems, e.~g., the bow-tie or the octahedron.

\section*{Acknowledgment}
H.-J.~S. would like to thank Martin Gemb\'e, Ciar\'an Hickey, Yasir Iqbal, Johannes Richter, and Simon Trebst
for the discussions on the square Kagome lattice that gave rise to the idea for this paper.

\appendix

\section{Comparison to other results}\label{sec:CO}

Using transfer matrix methods, a series representation of the partition function of the AF Heisenberg $N$-ring has been derived \cite{J67}
that reads
\begin{equation}\label{joyce}
 Z_N(\beta)=\sum_{\ell=0}^{\infty}(2\ell+1)\,\left(\sqrt{\frac{\pi }{2 \beta}}\, I_{\ell+\frac{1}{2}}(\beta )\right)^N
 \;.
\end{equation}
This series can be re-written using the definition of the modified spherical Bessel function
$i_\ell(z):=\sqrt{\frac{\pi }{2 z }} I_{\ell+\frac{1}{2}}(z)$, see \cite{NIST21}, 10.47.7..
For $N=2,3,4$ there exist independent expressions for $Z_N(\beta)$.
Let us start with $N=2$.

According to \cite{MSL99}, eq.(14) and setting $J_c=2$, we have
\begin{equation}\label{Z2}
 Z_2(\beta)=\frac{\sinh (2 \beta )}{2 \beta }
 \;.
\end{equation}
This result also follows from \cite{CLAL99}, eq.~(37),
in the limit of vanishing magnetic field and leads to the identity
\begin{equation}\label{joyce2}
\sum_{\ell=0}^{\infty}(2\ell+1) I_{\ell+\frac{1}{2}}(\beta )^2=
\frac{\sinh (2 \beta )}{\pi }
\;.
\end{equation}
The latter identity can also be obtained by multiplying the recurrence relation
\begin{equation}\label{modBrec}
 \frac{2\ell+1}{\beta}\,i_\ell(\beta)=i_{\ell-1}(\beta)-i_{\ell+1}(\beta)
 \;,
\end{equation}
see \cite{Detal89}, eq.~(37), with $i_\ell(\beta)$ and summing over $\ell=0,\ldots,\infty$.

\begin{figure}[htp]
\centering
\includegraphics[width=0.7\linewidth]{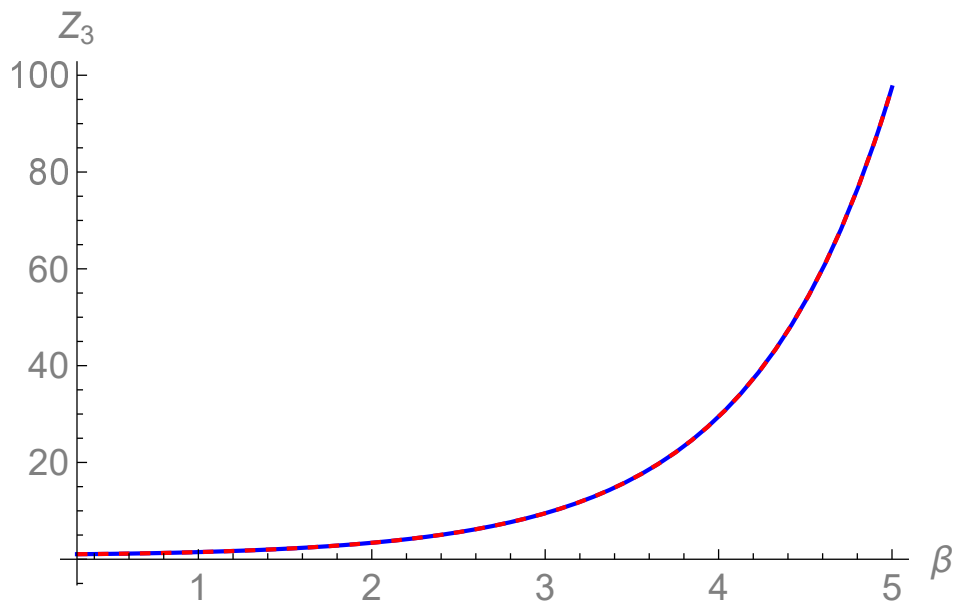}
\caption{The partition function $Z_3(\beta)$ of the spin triangle, calculated according to (\ref{Z3JM}) (blue curve), and
according to (\ref{Z3C}) (red dashed curve).
}
\label{FIGz3}
\end{figure}

Curiously, the analog procedure does not work for $N=3$. We adopt the explicit form of $Z_3$ from \cite{CLAL99}, eq.~(23),
up to the factor $(4\pi)^3$,
\begin{equation}\label{Z3C}
Z_3(\beta)= \frac{ e^{3 \beta /2}}{4 \beta^{3/2}}\,{\sqrt{\frac{\pi }{2}} \left(3 \,\text{erf}
\left(\sqrt{\frac{{\beta}}{2}}\right)-
\text{erf}\left(3 \sqrt{\frac{\beta}{2}}\right)\right)}
   \;.
\end{equation}
This equation has been independently checked using the methods of this paper. But (\ref{Z3C}) differs numerically from (\ref{joyce})
for $N=3$. Instead we found the following identity
\begin{equation}\label{Z3JM}
Z_3(\beta)= -\sum_{\ell=0}^{\infty } (2\ell+1) \left(\sqrt{-\frac{\pi }{2 \beta }} I_{\ell+\frac{1}{2}}(-\beta)\right){}^3
   \;,
\end{equation}
see Figure \ref{FIGz3}.
When evaluating the modified spherical Bessel function for negative real arguments
we have to consider that this function has a cut on the negative real axis, cp.~\cite{NIST21}, \S 10.25.
In (\ref{Z3JM}) we have used the convention, also adopted by MATHEMATICA, that $i_\ell(z)$ will be an odd function for even $\ell$ and vice versa.

Now we pass to $N=4$. Equating (\ref{joyce}) for $N=4$ with our expression for $Z_4(\beta)$ and multiplying with $4\beta^2$ we obtain the following identity
\begin{equation}\label{joyce4}
\sum_{\ell=0}^{\infty}(2\ell+1) I_{\ell+\frac{1}{2}}(\beta ){}^4=
\frac{1}{\pi ^2 }\left(\text{Chi}(4\beta)-\log (4 \beta )-\gamma \right)
\;.
\end{equation}

For the sake of completeness we add the analogous identity for $N=1$ which follows from \cite{AS72}, 10.2.26, and $\theta=0$,
but has no physical interpretation in terms of partition functions for spin systems:
\begin{equation}\label{joyce1}
\sum_{\ell=0}^{\infty}(2\ell+1) I_{\ell+\frac{1}{2}}(\beta)=
\sqrt{\frac{2\beta}{\pi}}\,\exp \beta
\;.
\end{equation}
This identity can also be derived by summing (\ref{modBrec}) over $\ell=0,\ldots,\infty$.\\

All identities mentioned here have been numerically tested.


\end{document}